\documentstyle[12pt]{article}
\title{\nopagebreak
\begin{flushright}
\tenrm UCTP103.98
\end{flushright}\vskip0.3in
\nopagebreak
\large \bf Supermultiplets of AdS Black Holes \\
in 2+1 Dimensions}
\author{Sharmanthie Fernando\thanks{email address:
fernando@physung.phy.uc.edu} and Freydoon Mansouri\thanks{email
address: Mansouri@uc.edu} \\
\it \small \it Physics Department, University of Cincinnati,
Cincinnati, OH 45221}
\date{}

\begin{document}
\maketitle

\begin{abstract}

We construct super AdS black holes in 2+1 dimensions
in terms of Chern
Simons gauge theory of $N=(1,1)$ super AdS group coupled to a
(super)source. We take the source to be a super AdS state
specified
by its Casimir invariants. We show that the corresponding space-
time is a supermultiplet of AdS space-times related to each other
by supersymmetry transformations. We give explicit expressions
for the masses and the angular momenta of the black holes in a
supermultiplet. With one exception, for $N=(1,1)$ one pair of
extremal black holes can be accommodated
in such all-black hole supermultiplets. The requirement that the
source be a unitary representation leads to a discrete tower of
excited states which provide a microscopic model for the super
black hole.
\end{abstract}

A conventional method of searching for signs of supersymmetry in
black hole solutions is to look for Killing spinors. Many
works along these lines already exist in the
literature. We cite a representative few
here~\cite{rone}-~\cite{rnine}, from which more references can be
traced. 
One way to see whether a given black hole solution admits Killing
spinors is identify it with the bosonic part
of an appropriate
supergravity theory~\cite{rone}-~\cite{rnine}. Then by requiring
that the fermions as well
as their variations vanish, one arrives at Killing spinor
equation(s). The asymptotic supersymmetries depend on the number
of non-trivial solutions of these equations consistent with the
black hole topology. For example, in asymptotically flat
space-times, a
typical supermultiplet consists of a black hole and a number of
ordinary particles all with the same mass. In contrast to the
familiar situation in particle physics, where we have
Supermultiplets consisting of particles only, in this approach
there is no systematic way of looking for
supermultiplets consisting of black holes only. The main purpose
of the present work is to show that in $2+1$
dimensions it is possible to construct a theory which permits
macroscopic solutions consisting of all AdS black hole
supermultiplets~\cite{rten}. It
involves the Chern Simons gauge theory of the (1,1) super AdS
group~\cite{releven,rtwelve} coupled to a super AdS state
(source). As we shall see below, to be able to accommodate the
structure of the solution which emerges from such a theory it
becomes necessary to broaden the standard notions of classical
geometries
to include some quantum mechanical elements.

The theory which we will describe below is the supersymmetric
generalization of a theory~\cite{rthirteen} which has been used
recently to provide both a microscopic and a macroscopic
description of the BTZ black hole~\cite{rfourteen}. Since the
concepts used in reference~\cite{rthirteen}, in which the gauge
group is the
AdS group, are essential for the understanding of the present
work, we begin with a brief summary of the results of that work.
This will also allow us to establish our notation and to describe
our unconventional use of Chern Simons theory. In the same sense
as in reference~\cite{rthirteen},
we take the Chern
Simons theory to be defined on a manifold $M$ with
topology $R \times \Sigma$, where $\Sigma$ is a two dimensional
space. Moreover, we consider the theory to be an explicit
realization of the Mach
Principle, so that in the absence of sources the field strengths
vanish and the topology is trivial (no punctures). In this way,
we associate non-trivial topologies to the presence of
sources~\cite{rfifteen,rsixteen}. We then show that the
information encoded in such a theory identifies the
physical (metrical) space-time as the space $M_q$ of $0+1$
dimensional fields which represent the source. Taking the source
to be a
super AdS state, we find
that the emerging 
physical space-time has a multilayered structure, distinct from
the manifold $M$ over which the gauge theory is defined. The
layers are connected to each other by
supersymmetry transformations. As a result,
for $N = (1,1)$ and appropriate ranges of Casimir invariants of
the source state, the physical space-
time becomes a super AdS (super)multiplet consisting of four AdS
black holes.

Let us now consider the details. It will be recalled that
the anti-de Sitter space in 2+1 dimensions can be viewed  as a
subspace of a
flat 4-dimensional space with the line element
\begin{equation}
ds^2 = dX_AdX^A = dX_0^2 - dX_1^2 -dX_2^2 + dX_3^2 \end{equation}
It is determined by the constraint
\begin{equation}
(X_0)^2 - (X_1)^2 - (X_2)^2 + (X_3)^2 = l^{2}\end{equation}
where $l$ is a real constant . The set of transformations which
leave the line element invariant
form the anti-de Sitter group $SO(2,2)$ which is locally
isomorphic
to $SL(2,R) \times SL(2,R)$. From
here on by
anti-de Sitter group we shall mean its universal covering group.

With $a= 0,1,2,$ we can write the AdS algebra in two
convenient forms~\cite{rthirteen}:
\begin{eqnarray}
J_a  = J^+_a +J^-_a  \nonumber \\
l\Pi^a = J^{+a} - J^{-a}  \end{eqnarray}
Setting
\begin{equation} 
\epsilon^{012} = 1 ; \hspace{1.cm} \eta^{ab} = (1, -1,
-1)\end{equation}
the commutation relations in, say, $J_a^{\pm}$ basis will take
the
form,
\begin{equation}
\left[J_a^{\pm}, J_b^+\right] = -i\epsilon_{ab}^c
J^{\pm}_{c};\;\;\;\;\;
\left[J_a^+, J_b^-\right] = 0 \end{equation}
The Casimir operators are then given by
\begin{equation}
j_{\pm}^2 = \eta^{ab} J^{\pm}_a J^{\pm}_b
\end{equation}
Alternatively, we can take a combination of these with
eigenvalues corresponding to the parameters of the BTZ solution:
\begin{eqnarray}
M = l^2 (\Pi^a\Pi_a + l^{-2}J^aJ_a) = 2(j_{+}^2 + j_{-}^2)
\nonumber \\
J/l = 2l\Pi_aJ^a = 2( j_+^2 - j_-^2)\;\;\;\;\;
\end{eqnarray}
Unless stated otherwise, we will use the same symbols for
operators and their eigenvalues.
As explained elsewhere~\cite{rthirteen}, an irreducible
representation associated with an
 AdS black hole can be labeled either with the pair $(M,J)$ or
the pair $(j_+,j_-)$. It is also possible to label these
representations by eigenvalues which are proportional to the
Horizon radii $r_{\pm}$ of the AdS black hole~\cite{rthirteen}.

To write down the Chern Simons action,
we begin by expressing the connection in $SL(2,R) \times SL(2,R)$
basis
\begin{equation}
A_{\mu} = \omega ^a_{\mu} J_a +
e^a_{\mu} \Pi_a
= A^{+a}_{\mu} J_a^+ + A_{\mu}^{-a} J_a^- \end{equation}
where 
\begin{equation}
A^{\pm a}_{\mu} = \omega ^a_{\mu} \pm l^{-1} e^a_{\mu}
\end{equation}
Eq. (9) should be viewed as definitions of $e$ and
$\omega$ in terms of the two $SL(2,R)$ connections. The covariant
derivative will have the form
\begin{equation}
D_{\mu} = \partial_{\mu} - iA_{\mu}= \partial_{\mu}
-iA^{+a}_{\mu} J_a^+ -i A_{\mu}^{-a} J_a^- \end{equation}
Then the components of the field strength are given by
\begin{equation}
[D_{\mu}, D_{\nu}] = -iF^{+a}_{\mu \nu} J^+_a - iF^{-a}_{\mu
\nu} J^-_a 
 = -iF^{+}_{\mu \nu}[A^+] -  iF^{-}_{\mu \nu}[A^-] \end{equation}
For a simple or a semi-simple group, the Chern Simons action has
the form
\begin{equation}
I_{cs} = \frac{1}{4\pi}Tr \int_M A \wedge \left( dA + \frac{2}{3}
A \wedge A\right) \end{equation}
where Tr stands for trace and
\begin{equation}
A = A_{\mu} dx^{\mu} =  A^+ + A^- \end{equation} 
So, The Chern Simons action with $SL(2,R)\times SL(2,R)$ gauge
group will take the form 
\begin{equation}
I_{cs} = \frac{1}{4\pi}Tr \int_M \left[\frac{1}{a_{+}}A^+ \wedge
\left( dA^+ +
\frac{2}{3}
A^+ \wedge A^+ \right) + \frac{1}{a_{-}}A^- \wedge \left( dA^- +
\frac{2}{3}
A^- \wedge A^- \right) \right] \end{equation}
Here the quantities $a_{\pm}$ are, in general, arbitrary
coefficients, reflecting the semisimplicity of the gauge group.
Up to an overall normalization, only their ratio
is significant. In the presence of a source (or of sources) in
$M$, any
\'a priori choice of the coefficients $a_{\pm}$ reduces the class
of allowed holonomies~\cite{rthirteen}. We will, therefore, keep
the
coefficients $a_{\pm}$ as free parameters in the sequel. 

As stated above, the manifold $M$ has the topology $R \times
\Sigma$ with R representing $x^{0}$. Then subject to the
constraints
\begin{equation}
F^{\pm}_a[A^{\pm}] =\frac {1}{2}  \eta_{ab} \epsilon^{ij}
(\partial_i
A_j^{\pm\;b} - \partial_j A_i^{\pm\;b} + \epsilon^b_{\;cd}
A_i^{\pm\;c}
A_j^{\pm\;d}) = 0 \end{equation}
the Chern Simons action for $SO(2,2)$ will take the form
\begin{eqnarray}
2\pi I_{cs} = \frac{1}{a_+} \int_R dx^{0}  \int_{\Sigma}
d^2x\left(- 
\epsilon^{ij}\eta_{ab} A^{+a}_i \partial_0 A^{+b}_j +   A^{+a}_ 0
F^+_a \right)\nonumber \\
+ \frac{1}{a_-} \int_R dx^{0}  \int_{\Sigma} d^2x\left(- 
\epsilon^{ij}\eta_{ab} A^{-a}_i \partial_0 A^{-b}_j +  A^{-a}_0
F_a^- \right)\end{eqnarray}
where $i,j=1,2$.

To introduce interactions, we follow an approach which has been
successful in coupling
sources to Poincar\'e and super Poincar\'e Chern Simons
theories~\cite{rfifteen,rsixteen} and take a source for the
present problem to
be an irreducible representation
of anti-de-Sitter group characterized by Casimir invariants $M$
and $J$. 
Within the representation, the states are further  specified by
the phase space
variables of the source $\Pi^A$ and $q^A$, $A= 0,1,2,3$, subject
to anti-de Sitter  constraints. The relevant irreducible
representations of the AdS group have been discussed in
reference~\cite{rthirteen}. Here we note that to allow for
the possibility of quantizing the Chern Simons theory
consistently, we must require that our sources be represented by
unitary representations of AdS group. 
Since the AdS group in $2+1$ dimensions can be represented in the
$SL(2,R) \times SL(2,R)$ form, the unitary
representations of $SO(2,2)$ can be constructed from those of
$SL(2,R)$. The latter group has four series of unitary
representations all of which are infinite
dimensional~\cite{rseventeen}. Of these,
the relevant representations for our purposes turn out to be the
discrete series bounded from below~\cite{rthirteen}.
For this series, the states in
an irreducible representation of $SL(2,R)$ are specified by the
eigenvalues of its Casimir operator $j^2$ (see Eq. 6) and, e.g.,
the element $J_0$, where we have suppressed the superscripts
$\pm$ distinguishing our two $SL(2,R)$ 's. Thus, we
have
$$j^2 |F, m> = F (F - 1) |F, m>$$
$$J_0 |F, m> = (F + m) |F, m>$$
In these expressions
\begin{equation}
F = real \;\; number \geq 0; \;\;\;\;\; m
= 0, 1, 2,...
\end{equation}
So, for this series,
the eigenvalues of the Casimir
invariants of $SL(2,R) \times SL(2,R)$
can be written as, 
\begin{equation}
j^2_{\pm} = F^2_{\pm} - F_{\pm}
\end{equation}
It follows that the infinite set of states can, in a somewhat
redundant notation, be
specified as
\begin{equation}
 |j_{\pm}^2, F_{\pm}+ m_{\pm}>; \;\;\;\;\; m_{\pm} = 0, 1, 2, ...
\end{equation}
Clearly, the integers $m_{\pm}$ are not necessarily equal.
Using these states, we can construct the discrete series of the
unitary
representations of $SO(2,2)$. A typical state
will have the following labels:
\begin{equation}
| M, J > = | j_+^2, j_-^2, F_+ + m_+, F_- + m_- >
\end{equation}
To be able to identify the labels $M$ and $J$ with the
corresponding labels in the AdS black hole, we must require that
$F_{\pm} \geq 1$~\cite{rthirteen}. It would then follow that
$|J/l| \leq
|M|$, as required for having a black hole solution.

With this background, let us now consider the coupling of sources
to the AdS Chern Simons theory. It is given by
$$I_{s}  = \int_C d \tau\left[\Pi_A \partial_{\tau}q^A -
(A^{+a}J^+_a + A^{-a}J^-_a)+ \lambda\left( q^Aq_A - l^2
\right)\right]$$ 
\begin{equation}
+\int_C d\tau\left[ \lambda_+ \left( J^{+a} J^{+}_a -
l^2j_+^2\right)+
\lambda_-
\left(J^{-a} J_a^- - l^2j_-^2\right)\right]
\end{equation}
In this expression, $C$ is a path in $M$, $\tau$ is a  parameter
along $C$, and $J^{\pm}_a$ play the role of c-number
generalized
 angular momenta which transform in the same way as the
corresponding generators which label the source. The quantities
$\lambda$ and $\lambda_{\pm}$ are Lagrange multipliers.
The first constraint in this action ensures that $q^A(\tau)$
satisfy the AdS
constraint. As explained in previous
occasions~\cite{rthirteen,rfourteen,rfifteen}, it is
not the manifold $M$ over which
the gauge theory is
defined but the space of $q_A$'s which give rise to the classical
space-time. The last 
two constraints identify the source being coupled to the Chern
Simons theory as an
anti-de Sitter state with invariants $j_+$ and $j_-$. These
constraints are crucial in relating
the invariants of the source to the asymptotic observables of the
coupled theory via Wilson loops.

The total action for the theory is given by:
\begin{equation}
I = I_{cs} + I_{s} \end{equation}
It is easy to check that in this theory
the components of the field strength still vanish
everywhere except at the location of the sources.
So, the analog of Eq. 15 become
\begin{equation}
\epsilon^{ij} F^{\pm\;a}_{ij} =2\pi a_{\pm} J^{\pm\;a}
\delta^2(\vec{x},\vec{x_0})
\end{equation}
In particular, fixing the gauge so that $SO(2,2)$ symmetry
reduces to $SO(2) \times SO(2)$, we get
\begin{equation}
\epsilon^{ij} F^{\pm\;0}_{ij} = 2\pi a_{\pm} F_{\pm}
\delta^2(\vec{x},\vec{x^0})
\end{equation}
where $F_{\pm}$ are the invariant labels of the state as in Eq.
18. All other components of the field strength vanish. 
We thus
see
that because of the 
constraints appearing in the action given by Eq. 22, the strength
of the
sources corresponding to the maximal compact subgroup of the
gauge group
become related 
to their Casimir invariants. These invariants,
in turn, determine the asymptotic observables of the theory.
Since such 
observables must be gauge invariant, they are expressible in
terms of Wilson loops,
and a Wilson loop about our source can only depend on, e.g.,
$j_+$ and $j_-$.

From the data on the manifold $M$ given above, one can
determine the properties of the emerging space-time by solving Eq
24. The only
non-vanishing components of the gauge potential are given
by~\cite{rthirteen}
\begin{equation}
A^{\pm 0}_{\theta} = 2a_{\pm} F_{\pm}
\end{equation}
where $\theta$ is an angular variable. In particular, using Eq.
9, we can write~\cite{rthirteen}
\begin{equation}
\omega^0_{\theta}/l = (j_+ - j_-) = r_-/l
\end{equation}
\begin{equation}
e^0_{\theta} = (j_+ + j_-) = r_+/l
\end{equation}
Although these are components of a connection which is a pure
gauge, they give rise to non-trivial holonomies
around the source. More explicitly, we have
\begin{equation}
W[e] = exp^{ \oint_{\gamma} e^{0}_{\theta} \Pi_0}
\end{equation}
\begin{equation}
W[\omega] = exp^{ \oint_{\gamma} \omega^{0}_{\theta}J_0}
\end{equation}
Here, $\gamma$ is a loop around the source. 

The reduction from Eq. 23 to Eq. 24 with $SO(2) \times SO(2)$ as
left over symmetry relies on diagonalizing compact subgroup
generators. Although this can work for black hole solutions as
described in reference~\cite{rthirteen}, the natural left over
symmetry from the point of view of black holes is $SO(1,1) \times
SO(1,1)$. Starting from Eq. 23, one can carry out the gauge
fixing such that the analog of Eq. 24 will have this non-compact
symmetry~\cite{reighteen}. The analysis of the holonomies will go
through as in the case of the compact subgroup. The main
advantage in this case is that it is no longer necessary to carry
out a Wick rotation to make contact with the black hole
solution~\cite{rthirteen}.

Irrespectively of whether the left over symmetry were compact or
non-compact, it was shown in
reference [13] how these holonomies lead to a discrete
identification subgroup of $SO(2,2)$, which shows that the
manifold $M_q$ of the $0+1$ dimensional fields $q^A$ has all the
relevant features of the macroscopic AdS black hole solution. The
approach used in this reference was to improve and make precise
some of the previous attempts to obtain this kind of 
identifications~\cite{rnineteen,rfifteen,rsixteen,rtwenty}. As
we shall see below, the same holonomies, suitably interpreted,
will play a crucial role in establishing the space-time structure
of the supersymmetric theory discussed below.
To actually obtain the BTZ line element, one must
determine a parametrization of $q^A$'s consistent with the above
holonomy properties. This was carried out in
reference~\cite{rthirteen}, a typical parametrization for $r >
r_+$ being
\begin{eqnarray}
q^1 = f cos \left( \frac{r_-}{l}\phi - \frac{r_+t}{l^2}
\right)\;\;\;\;\;\;\nonumber \\
q^2 = f sin \left( \frac{r_-}{l}\phi - \frac{r_+t}{l^2}
\right)\;\;\;\;\;\;\nonumber \\
q^0 = \sqrt{ f^2 + l^2 } cos \left( \frac{r_+}{l}\phi -
\frac{r_-t}{l^2} \right)\nonumber \\
q^3 = \sqrt{ f^2 + l^2 } sin  \left( \frac{r_+}{l}\phi -
\frac{r_-t}{l^2} \right) \end{eqnarray}
where
\begin{equation}
 \frac{f^2}{l^2} =  \frac{ r^2 - r_+^2}{r_+^2 - r_-^2};
\hspace{0.5in} r > r_+ \end{equation}
the important point to note here is that the quantities $q^A$
carry the Casimir invariants
$(r_+,r_-)$ of the source state.

We now turn to the supersymmetric generalization of the Chern
Simons theory described above and show that the corresponding
macroscopic theory consists of a supermultiplet of ordinary
space-times and, as a special case, a supermultiplet consisting
of black holes only.
The simplest way of obtaining the supersymmetric extension of the
anti-de Sitter group is
to begin with the AdS group in its $SL(2,R) \times SL(2,R)$
basis. The $N=1$ supersymmetric form of each $SL(2,R)$ factor is
the supergroup $OSp(1|2;R)$. Thus, one arrives at the (1,1) form
of the $N=2$ super AdS group. Its algebra is given by
\begin{eqnarray}
[J_a^{\pm}, J_b^{\pm}] = -i\epsilon_{ab}^{\;\;\;c} J_c^{\pm};
\;\;\; 
[J^{\pm}_a,Q^{\pm}_{\alpha}] =
-\sigma^{a\;\beta}_{\alpha}Q^{\pm}_{\alpha}; \;\;\; 
\{Q^{\pm}_{\alpha}, Q^{\pm}_{\beta} \} =
-\sigma^{a}_{\alpha\beta}
J^{\pm}_a \;\;\;\;\;\nonumber \\ 
\{Q^+_{\alpha}, Q^-_{\beta}\} = 0;  \;\;\;\;\; [J^+,J^-]= 0
\;\;\;\;\;\;\;\;\;\;\;\;\;\;\;\;\;\;\;\;\;\;\;\;\;\end{eqnarray}
The Casimir invariants are given by
\begin{equation}
C_{\pm} = j_{\pm}^2 +
\epsilon^{\alpha\beta} Q^{\pm}_{\alpha} Q^{\pm}_{\beta}
\end{equation}
The spinor indices are raised and lowered by antisymmetric metric
$\epsilon^{\alpha \beta}$ defined by
$\epsilon^{12} = -\epsilon_{12} = 1 $. The matrices
$(\sigma^a)_{\alpha}^{\beta}$, $( a= 0,1,2)$, form a
representation of 
$SL(2,R)$ and satisfy the Clifford algebra
\begin{equation}
\{\sigma^a,\sigma^b\} = \frac{1}{2}\eta^{ab}\end{equation}
More explicitly, we can take them to be:
\begin{eqnarray}
\sigma^0 = \frac{1}{2} \left(\begin{array}{cc} 1 & 0 \\ 0 & -1
\end{array} \right) ;\;\;\;\; 
\sigma^1 = \frac{1}{2} \left(\begin{array}{cc}  0 & i \\ i & 0
\end{array} \right) ;\;\;\;\; 
\sigma^2 = \frac{1}{2}\left(\begin{array}{cc} 0 & 1 \\ -1 & 0
\end{array} \right)
\end{eqnarray}
It is important to note that the supersymmetry generators of
$OSp(1|2,R)$ do not commute with the Casimir invariant of its
$SL(2,R)$ subgroup. That is,
\begin{equation}
[j^2_{\pm}, Q_{\alpha}] \neq 0
\end{equation}
 
Since super AdS group is semi-simple, we can construct its
irreducible representations by first constructing the irreducible
representations of $OSp(1|2,R)$. Depending on which $OSp(1|2,R)$
we are considering, the states within any such supermultiplet are
the corresponding irreducible representations of 
$SL(2,R)$ Characterized by the Casimir invariants $j_+$ and 
$j_-$,
respectively. Based on the rationale given for the non-
supersymmetric case, the irreducible representations of interest
for the present case are those which can be obtained from the
unitary
discrete series of $SL(2,R)$ and which are bounded from below. To
construct the supermultiplet corresponding to,
say, the ``plus'' generators in Eq. 32, we can take the Clifford
vacuum state
$|\Omega^+>$ to be the $SL(2,R)$ state with the lowest eigenvalue
of $J_o^+$. In the notation of Eq. 19, this corresponds to an
$m=0$ state:
$$|F_+, m > = |F_+, m=0 > = |F_+ >$$
Then, the superpartner of this state, again with $m=0$, is the
state $|F_+ +1/2>$ obtained by the application of one of the
$Q$'s. The corresponding values of $j^2_+$ are $F_+(F_+ -1)$ and
$(F_+ +1/2)(F_+ -1/2)$, respectively. The supermultiplet for the
second
$OSp(1|2,R)$ can be constructed in a similar way.

We are now in a position to construct the (1,1) super AdS
supermultiplet as a direct product of the two $OSp(1|2,R)$
doublets. Altogether, there will be four states in the
supermultiplet. They will have the following labels:
\begin{equation}
|F_+,F_- >; \hspace{0.4cm} |F_+ + 1/2,F_- >;
\hspace{0.4cm} |F_+, F_- + 1/2>; \hspace{0.4cm}
|F_+ + 1/2, F_- + 1/2 > \end{equation}
From these, we can also obtain the  expressions for the
eigenvalues $(M,J)$ of various states within the supermultiplet:
\begin{eqnarray}
|M_1, J_1 > = |M, J > \nonumber \\
|M_2, J_2 > = |M + 2F_+ - 1/2, J + 2F_+ - 1/2 > \nonumber \\
|M_3, J_3 > = |M + 2F_- - 1/2, J - 2F_- + 1/2 > \nonumber \\
|M_4, J_4 > = |M + 2(F_+ + F_-) - 1, J + 2(F_+ - F_-) >
\end{eqnarray}
these states transform into one another under supersymmetry
transformations.

The Chern Simons action for simple and semisimple supergroups has
the same structure as that for Lie groups. The only difference is
that the trace operation is replaced by super trace (Str)
operation. So, in the $OSp(1|2,R)
\times OSp(1|2,R)$ basis the
Chern Simons action for the super AdS group has the same form as
that given by Eq. 14. But now the
expression for connection is given by
\begin{equation}
A^{\pm} = \left[ A^{\pm\;a}_{\mu} J^{\pm}_a +
\chi^{\pm\alpha}_{\mu}
Q^{\pm}_{\alpha}\right]dx^{\mu}\end{equation}
Just as in the non-supersymmetric case, to have a
nontrivial theory, we must couple sources to the Chern Simons
action. To do this in a gauge
invariant and locally supersymmetric fashion, we must take a
source to be an irreducible representation of the super AdS
group. As we saw above, such a supermultiplet
consists of four AdS states. To couple it to the gauge fields, we
must first generalize the canonical variables we used in the AdS
theory to their supersymmetric forms:
\begin{equation}
 \Pi_A \rightarrow (\Pi_A,\Pi_{\alpha}) \hspace{0.5in} q_A
\rightarrow (q_A,q_{\alpha})\end{equation}
Then, the source coupling can be written as
\begin{equation}
I_s = \int_C \left[\Pi_Adq^A + \Pi_{\alpha}dq^{\alpha} + (A^+ 
+ A^-)+ constraints \right] \end{equation}
where again $C$ is a path in $M$.
The constraints here include those discussed for the AdS group
and, in addition, those which relate the AdS labels of
the Clifford vacuum to the Casimir eigenvalues of the super AdS
group. The combined action
\begin{equation}
I = I_{cs} + I_s \end{equation}
leads to the constraint equations
\begin{equation}
\epsilon^{ij} F^{\pm\;a}_{ij} = 2\pi a_{\pm} J^{\pm\;} \delta^2
(\vec{x},\vec{x^0}); \hspace{0.5in} 
\epsilon^{ij} F^{\pm\alpha}_{ij} = 2\pi a_{\pm} 
Q^{\pm\alpha}\delta^2(\vec{x},\vec{x^0}) \end{equation}

Up to this point, everything proceeds in parallel with the non-
supersymmetric case. Differences begin to show up when one
attempts to solve these equations by
choosing a gauge again so that the gauge symmetry is reduced to
$SO(2) \times SO(2)$:
\begin{equation}
\epsilon^{ij}F^{\pm\;0} = 2\pi a_{\pm} J_0^{\pm}
\delta^2(\vec{x},\vec{x^0}) \end{equation}
Although this equation is identical in form to Eq. 24 for the
non-supersymmetric case, there is an essential difference in the
underlying physics. In the supersymmetric case, the
supermultiplet which we couple to the Chern Simons action consist
of four $SO(2,2)$ states with different values of $F_{\pm}$. As a
result, in the parallel transport of $q^A$ around a close path
analogous to the non-supersymmetric case, there will be four sets
of holonomies with different values of $(j_+, j_-)$ or,
equivalently, $(r_+, r_-)$. Moreover, in the non-supersymmetric
case, a single source with Casimir invariants $(r_+, r_-)$ or,
equivalently, $(M,J)$ will give rise to an AdS black
hole~\cite{rfourteen} for which the 
line element is characterized with the corresponding values of
$M$ and $J$:
\begin{equation}
ds^2= -(\frac{r^2}{l^2} -M + \frac{J^2}{4l^2}) dt^2 +
\frac{dr^2}{(\frac{r^2}{l^2} -M + \frac{J^2}{4r^2})} + r^2 (d\phi
- \frac{J^2}{2r^2}dt)^2 \end{equation}
In the supersymmetric case, the source is a supermultiplet in
which there are four states of differing  $(M,J)$ values.
Moreover, recall that
the explicit parametrization of $q^A$ given by Eq. 30, depends on
the Casimir invariants $(r_+,r_-)$ or, equivalently, on $(M,J)$.
Then, 
depending on which set $(M,J)$ that we choose, we will get a
different BTZ solution. Since $M$ and $J$ are not invariant under
supersymmetry transformations, these solutions are transformed
into each other under supersymmetry.
This makes it
impossible for a single c-number line element of the type given
by Eq. 45 to
correspond to all the AdS states of a supermultiplet.

The situation
here runs parallel to what was encountered in connection with
super Poincar\'e Chern Simons theory~\cite{rsixteen}. There it
was
pointed out that standard classical geometries were not capable
describing these
structures and that one must make use of {\it nonclassical
geometries}. Such
geometries can be based on three elements: 

{\bf 1}. An algebra such as a Lie algebra or a Lie superalgebra. 

{\bf 2}. A line element operator with values in this algebra. 

{\bf 3}. A Hilbert space on which the algebra acts linearly. 

For the problem at hand, the algebra of interest is the $N =
(1,1)$ super AdS algebra in $2+1$ dimensions. The corresponding
Hilbert space is the representation space of the superalgebra
given by Eq. 38. Then, instead of the BTZ line element given
above, we begin with a line element operator with values in the
$N = (1,1)$ superalgebra and assume that its diagonal elements
depend on the algebra only through the Casimir operators
$(\hat{M},\hat{J})$ of its $SO(2,2)$ subalgebra. The ``hats'' on
top of $M$ and $J$ are meant to distinguish the operators from
the corresponding eigenvalues. Thus, we have

$$ds^2 = ds^2(\hat{M}, \hat{J})$$
The matrix element of this operator for each state of the
supermultiplet will produce a c-number line element:
\begin{equation}
<M_k, J_k| ds^2 (\hat{M}, \hat{J}) |M_k, J_k> = ds^2(M_k,J_k)
\end{equation}
In other words, for each state of the supermultiplet, the
nonclassical geometry produces a layer of classical space-time.
The number of the layers is equal to the dimension of the
supermultiplet. Supersymmetry transformations act as messengers
linking different layers of this multilayered space-time. For
consistency, we must also interpret the quantities $J_0^{\pm}$ in
Eq. 44 as operators. Acting on different states of the
supermultiplet, they will give the corresponding $F_{\pm}$
eigenvalues. There will therefore be not one set but four sets of
holonomies W[e] and W[$\omega$]. Each set will produce the
discrete identification subgroup in the corresponding layer of
space-time. 

Consider, next, the conditions under which every layer of the
supermultiplet corresponds to an AdS black hole. For this to be
true, we must have
$$M_k \geq 0;\;\;\;\;\; |J_k| \leq lM_k$$
This, in turn implies that
$$F_+ \geq F_- \geq 1$$
In the notation of Eq. 38, for $|J| = lM$, two layers of the
supermultiplet become extreme AdS black holes. The only exception
is in the limiting case when $M=J=0$, in which case there will be
three extremal black holes in the supermultiplet. It is also
interesting to note that for an appropriate choice of $M$ and $J$
or, equivalently, $F_+$ and $F_-$, the same supersource which
generates a black hole in one layer can generate a naked
singularity in another. 

\bigskip
This work was supported, in part by the Department of Energy
under the contract number DOE-FGO2-84ER40153. The hospitality of
Aspen Center for Physics, where part of this work was carried
out, is gratefully acknowledged.

\end{document}